\documentclass[journal]{IEEEtran}

\usepackage{times}
\usepackage{mathrsfs}
\usepackage{epsfig}
\usepackage{graphicx}
\usepackage{amsmath}
\usepackage{amssymb}
\usepackage{graphicx}
\usepackage{cite}
\usepackage{enumerate}
\usepackage{cases}
\usepackage{multirow}
\usepackage{verbatim}
\usepackage{amssymb}
\usepackage{CJK}
\usepackage{algorithm}
\usepackage{algorithmicx}
\usepackage{algpseudocode}
\usepackage{color}
\usepackage{bm}
\usepackage{booktabs}
\usepackage[colorlinks,linkcolor=blue,citecolor=blue]{hyperref}
\usepackage{amssymb}
\usepackage[english]{babel}
\usepackage[utf8]{inputenc}
\usepackage[T1]{fontenc}
\usepackage{soul}

\newcommand{\etal}{\textit{et al}.}
\newcommand{\ie}{\textit{i}.\textit{e}.}
\newcommand{\eg}{\textit{e}.\textit{g}.}
\newcommand{\etc}{\textit{etc}}

\begin{document}

\title{Boundary Guided Semantic Learning for Real-time COVID-19 Lung Infection Segmentation System}

\author
{Runmin Cong,~\IEEEmembership{Member,~IEEE,} 
Yumo Zhang, 
Ning Yang,
Haisheng Li,
Xueqi Zhang, 
Ruochen Li, \\
Zewen Chen, 
Yao Zhao,~\IEEEmembership{Senior Member,~IEEE,}
and Sam Kwong,~\IEEEmembership{Fellow,~IEEE}

\thanks{Runmin Cong is with the Institute of Information Science, Beijing Jiaotong University, Beijing 100044, China, and also with Beijing Key Laboratory of Big Data Technology for Food Safety, Beijing Technology and Business University, Beijing 100048, China, also with the Beijing Key Laboratory of Advanced Information Science and Network Technology, Beijing 100044, China, and also with the Department of Computer Science, City University of Hong Kong, Hong Kong SAR, China (e-mail: rmcong@bjtu.edu.cn).}
\thanks{Yumo Zhang, Ning Yang, Xueqi Zhang, Ruochen Li, Zewen Chen and Yao Zhao are with the Institute of Information Science, Beijing Jiaotong University, Beijing 100044, China, and also with the Beijing Key Laboratory of Advanced Information Science and Network Technology, Beijing 100044, China (e-mail: yumozhang@bjtu.edu.cn; ningyang@bjtu.edu.cn; xueqizhang@bjtu.edu.cn; ruochenli@bjtu.edu.cn; zewenchen@bjtu.edu.cn; yzhao@bjtu.edu.cn).}
\thanks{Haisheng Li is with the Beijing Key Laboratory of Big Data Technology for Food Safety, Beijing Technology and Business University, Beijing 100048, China (e-mail: li\_haisheng@163.com).}
\thanks{Sam Kwong is with the Department of Computer Science, City University of Hong Kong, Hong Kong SAR, China, and also with the City University of Hong Kong Shenzhen Research Institute, Shenzhen 51800, China (e-mail: cssamk@cityu.edu.hk).}
}

\markboth{IEEE Transactions on Consumer Electronics}
{Shell \MakeLowercase{\textit{et al.}}: Bare Demo of IEEEtran.cls for IEEE Journals}
\maketitle

\begin{abstract}
The coronavirus disease 2019 (COVID-19) continues to have a negative impact on healthcare systems around the world, though the vaccines have been developed and national vaccination coverage rate is steadily increasing. At the current stage, automatically segmenting the lung infection area from CT images is essential for the diagnosis and treatment of COVID-19. Thanks to the development of deep learning technology, some deep learning solutions for lung infection segmentation have been proposed. However, due to the scattered distribution, complex background interference and blurred boundaries, the accuracy and completeness of the existing models are still unsatisfactory. To this end, we propose a boundary guided semantic learning network (BSNet) in this paper. On the one hand, the dual-branch semantic enhancement module that combines the top-level semantic preservation and progressive semantic integration is designed to model the complementary relationship between different high-level features, thereby promoting the generation of more complete segmentation results. On the other hand, the mirror-symmetric boundary guidance module is proposed to accurately detect the boundaries of the lesion regions in a mirror-symmetric way. Experiments on the publicly available dataset demonstrate that our BSNet outperforms the existing state-of-the-art competitors and achieves a real-time inference speed of 44 FPS. The code and results of our BSNet can be found from the link of \url{https://github.com/rmcong/BSNet}.
\end{abstract}


\IEEEpeerreviewmaketitle

\section{Introduction} \label{sec1}
\IEEEPARstart{T}{he} outbreak of coronavirus disease (COVID-19) has created a global, disruptive, long-lasting, and unprecedented public health crisis. More than 452.22 million people have been reported to be infected by the COVID-19 globally and 6.02 million people have died, according to Reuters statistics. Moreover, the virus is constantly mutating (\eg, delta, omicron), and many new virus variants have appeared. 
To complement the Reverse Transcription-Polymerase Chain Reaction (RT-PCR) testing, chest X-ray (CXR) and computed tomography (CT) have been widely used as the auxiliary screening tools for COVID-19 infection, which can be further used to classify the confirmed cases and formulate corresponding treatment methods.
In this paper, we utilize chest CT images as processing data to design an algorithm for automatically segmenting lung infections in COVID-19 cases.                      

\begin{figure}[!t]
\centerline{\includegraphics[width=0.5\textwidth]{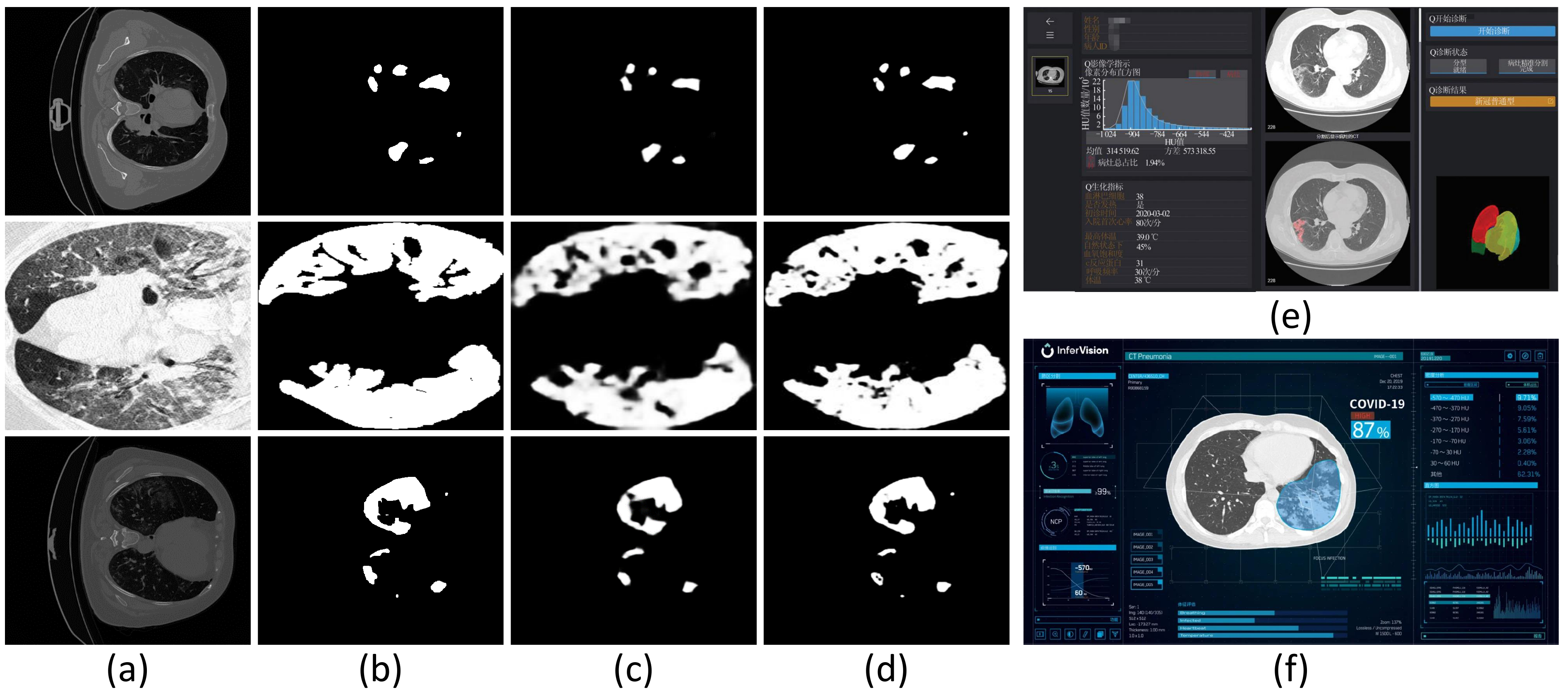}}
\caption{Visual examples of COVID-19 infection segmentation from CT images. (a) CT images. (b) Ground Truth. (c) Inf-Net. (d) Proposed BSNet. (e)-(f) COVID-19 intelligent diagnosis system designed by Wuhan university and InferVision, respectively.}
\label{fig1}
\end{figure}

Thanks to the powerful feature representation capabilities of deep learning, it has been widely used to address the computer vision task, such as enhancement \cite{crmunderwater,crmDBLP:journals/spl/HuJCGS21,crmunderwater21,crmCVPR20}, detection \cite{rrnet,crmglnet,crmdafnet,dpanet,crmCoADNet,crm21mm,crmACMMM20-1,crm-nc,crmDBLP:journals/tmm/MaoJCGSK22,crmDBLP:journals/tip/WenYZCSZZBD21}, super-resolution \cite{crmSRInpaintor,crmblindSR22,crmbridgenet,crmdsrCVPR21,crmdsr2019tip,crmijcai}, and medical image processing including lung nodules segmentation \cite{intro-4wang2017central}, brain and brain-tumor segmentation \cite{intro-5cherukuri2017learning}, polyp segmentation \cite{crmpolyp}, brain image synthesis \cite{huang2020mcmt}, retinal image non-uniform illumination removal \cite{li2020nui} \etc. For each different task, due to the differences in imaging equipment and disease characteristics, different segmentation models need to be designed separately. 
As far as COVID-19 diagnosis is concerned, a number of algorithms based on deep learning for CXR and CT images  have been proposed \cite{r17,infnet,COPLENet,r20,AnamNet,wu2021jcs,wang2021focus,yan2021covid,kitrungrotsakul2021attention,huang2022multi,song2021augmented,wang2020end}.
Among them, CT image is more widely used in clinical practice due to its higher sensitivity and clarity, such as COVID-19 classification and segmentation task. 
The COVID-19 lung infection segmentation from CT images aims to locate the infected regions and generate a pixel-wise segmentation mask. As can be seen from the examples given in Figure \ref{fig1}, this is a very challenging task, mainly manifested in: 

On the one hand, missing and incomplete detection of infection regions are common problems in the existing methods. By observing all images in Figure \ref{fig1}, infection regions often not appear concentrated, but scattered in multiple locations of the image, leading to a missing and incomplete detection.     
Meanwhile, when the patients' lungs have overwhelming infection, incomplete detection may be encountered, such as the third row of Figure \ref{fig1}.   
In fact, these scattered infection regions or internal larger range infection regions are still correlated in semantic attributes. Regarding this issue, we try to utilize the wealth of semantic information available at high-level encoder features to guide the feature learning in the decoder stage. Therefore, we propose a Dual-branch Semantic Enhancement (DSE) module to aggregate high-level encoder features, thereby modeling the global relation of different regions or different parts of regions. In addition, structure of human body causes backgrounds of the lung CT image (non-infected regions) to be complex, and thus the background interference is detrimental to precise targeting. Our DSE module can also benefit for suppressing complex backgrounds through semantic and category attributes.

On the other hand, the detailed boundaries of infected areas are not sharp and clear enough. As a tool of assistant COVID-19 diagnosis, boundary information plays an important role. The smooth boundaries may not have a positive effect on doctors' diagnosis \cite{2018Unsupervised}, such as the third row of Figure \ref{fig1}. As we all know, the low-level features have higher spatial resolution and more detailed boundaries, which can supplement the decoding process to achieve boundary guidance. However, directly transmitting the coarse low-level features may cause additional redundancy interference. Considering better suppression of unimportant features, we propose a Mirror-symmetric Boundary Guidance (MBG) module that can purify the features learned from the encoder and obtain more discriminative infection-related features.

In addition to the technical issues involved in model design, as described in recent and important review papers \cite{roberts2021common, naude2020artificial, malik2021artificial}, some common-sense pitfalls and biases are waiting to be solved, including the data used for model development, the evaluation and reproducibility of designed model. We also step up efforts to these three areas, specifically as follows: 

(1) Data. As described in \cite{roberts2021common}, using a public dataset alone without additional new data may lead to community-wide overfitting on this dataset. Therefore, in order to ensure the generalization ability of the proposed model, we merge the two publicly available CT-based segmentation datasets to obtain 1018 CT images, which are further divided into 718 training images and 300 testing images. In this way, it also can avoid selection bias caused by COVID-19 images come from the same place. Furthermore, data augmentation is used during the training phase to alleviate the data shortage problem.

(2) Evaluation. For more comprehensive quantitative evaluation, we use six metrics, including Dice Similarity Coefficient, Sensitivity, Precision, Structure Measure, Enhance-alignment Measure, and Mean Absolute Error. These indicators are measured from multiple aspects such as segmentation accuracy, completeness, structural representation ability, \etc. It is evident that our model achieves competitive performance across different metrics, indicating that the results produced by our model are validated and reliable. 

(3) Reproducibility. In order to describe the details of our network structure succinctly and clearly, we provide a table of convolution parameters to list the details of each convolution block in the method introduction. In addition, as suggested in \cite{roberts2021common}, the image resizing, cropping, and normalization are used before model input, and more training details (\eg, number of epochs, learning rate, and the optimizer) are also provided. 

In summary, an end-to-end COVID-19 infection segmentation model is proposed, which focuses on semantic relation modeling and boundary details guidance. The good portability of our proposed method enable it can be easily transplanted to the existing intelligent diagnosis system such as the two intelligent diagnosis systems shown in Figure \ref{fig1}. On this basis, the quality of the infection segmentation results can be improved by updating the algorithm model without changing the hardware, thereby realizing the integrated application across different fields. The major contributions are summarized as follows.    

\begin{itemize}
	\item[1)]
	We propose an end-to-end boundary guided semantic learning method for accurate and real-time COVID-19 lung infection segmentation, which can be easily transplanted to existing COVID‑19 intelligent diagnosis system for algorithmic model upgrades. Our work belongs to the research of the underlying algorithm framework in the field of consumer electronics.
	\end{itemize}
\begin{itemize}
\item[2)]
	We design a DSE module to aggregate high-level features in a complementary dual-branch strategy, including the top-level semantic preservation and the progressive semantic integration, thereby modeling the semantic relations and forcing the generation of complete infection area segmentation.
\end{itemize}
\begin{itemize}
	\item[3)]
	We propose a MBG module to introduce the low-level boundary information in the feature decoding stage with a mirror-symmetric structure, which can ensure the complementarity and sufficiency of feature learning.
\end{itemize}
\begin{itemize}
	\item[4)]
	Comparing the proposed method with eleven state-of-the-art approaches, our method achieves the superior performance under six evaluation metrics. Besides, the model has a real-time inference speed of 44 FPS.
\end{itemize}


\begin{figure*}[!t]
\centerline{\includegraphics[width=1\textwidth]{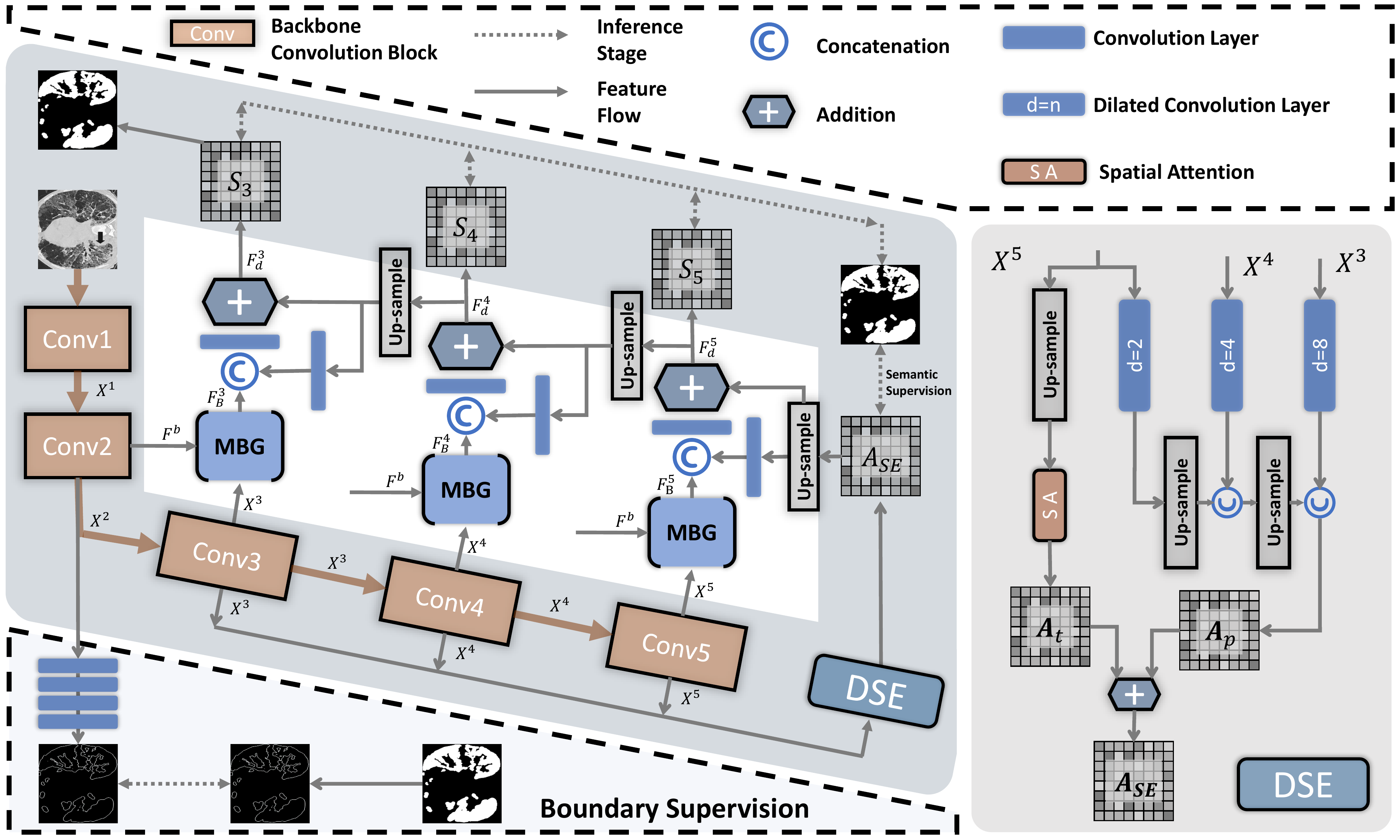}}
\caption{Illustration of the overall framework of proposed network. The input CT image is first fed into the backbone extractor to generate five multi-level features. The Dual-branch Semantic Enhancement (DSE) module is to aggregate high-level features, thereby generating a semantic attention mask to decoder. The features of last three stages are embedded with boundary features from the second stage by Mirror-symmetric Boundary Guidance (MBG) module. Outputs of MBG are combined with previous-stage features or semantic attention mask to produce three prediction maps, in which $S_3$ is the final result.}
\label{fig2}
\end{figure*}

\section{Related Work} \label{sec2}
\subsection{Medical Image Segmentation}
Convolutional neural networks became a popular machine learning algorithm for automated medical image analysis \cite{shen2017deep, hassantabar2021coviddeep, sayeed2019neuro, joshi2020iglu} due to the breakthrough of deep learning for computer vision. 
Most of the medical image segmentation methods are based on U-Net \cite{unet} structure or its modifications, such as UNet++ \cite{unet++}, Attention\_UNet \cite{attentionunet}, ResNet34\_UNet \cite{unet}.


Currently, thin-slice chest scans have become indispensable in thoracic radiology, but the huge amount of data also substantially increases the load of radiologists. As a result, automated chest CT image segmentation has become a popular auxiliary technique for lung disease diagnosis.
For example, Shen \etal \cite{shen2015automated} designed an automated lung segmentation system to boost the segmentation accuracy by utilizing the bidirectional chain code. 
Compared to classical machine learning methods, the deep learning algorithms can extract features from the perspective of semantic relation, which helps to segment nodule regions accurately from the similar visual background. 
Wang \etal \cite{intro-4wang2017central} proposed a central focused CNN to segment lung nodules in chest CT images, and designed a weighted sampling to facilitate the model training. 
Jin \etal \cite{jin2018ct} utilized GAN-synthesized data to improve the training of a discriminative model for pathological lung segmentation. 

\subsection{COVID-19 Lung Infection Segmentation}

So far, many COVID-19 lung infection segmentation methods from CT Images based on deep learning have been proposed, and promising performance has been obtained \cite{r17,infnet,COPLENet,r20,AnamNet,wu2021jcs,wang2021focus,yan2021covid,kitrungrotsakul2021attention,crm2022tim}. 
Zhou \etal \cite{r17} integrated the spatial and channel attention mechanisms to automatically segment the infection area. 
Fan \etal \cite{infnet} presented the parallel partial decoder, reverse attention, and edge-attention specifically for COVID-19 to improve the performance, and also provided a semi-supervised framework to alleviate the shortage of labeled data. 
Wang \etal \cite{COPLENet} proposed a noise-robust learning framework based on self-ensembling of convolutional neural network. 
Paluru \etal \cite{AnamNet} developed an anamorphic depth embedding-based lightweight convolutional neural network to segment anomalies in COVID-19 chest CT slices. 

Wu \etal \cite{wu2021jcs} proposed a novel joint classification and segmentation system to perform real-time and explainable chest CT diagnosis. 
Because the high intra-class variation and inter-class indistinction in COVID-19 infection appearance, Wang \etal \cite{wang2021focus} employed the autofocus and panorama modules for integrating the peer- and cross-level contexts. 
Yan \etal \cite{yan2021covid} introduced a feature variation block which adaptively adjusts the global properties of the features for segmenting COVID-19 infection. 
Kitrungrotsakul \etal \cite{kitrungrotsakul2021attention} proposed an interactive attention refinement network and an automatic seed point generation technique for the training.

\section{Proposed Method} \label{sec3}

\subsection{Overview}
As illustrated in Figure \ref{fig2}, an end-to-end network named BSNet is proposed for COVID-19 lung infection segmentation in CT images, following an encoder-decoder architecture. The overall framework can be divided into the encoder stage and decoder stage. Specifically, our backbone encoder extractor consists of five sequentially-stacked convolutional blocks, thereby obtaining multi-level features $\{X^1,X^2,X^3,X^4,X^5\}$. For clarity, we list the details of each convolution block in Table \ref{table:0}, where Res2Net \cite{Res2Net} is used as our backbone feature extractor, and RFB \cite{RFB} is a receptive field block used to enhance features learned from Res2Net.

\begin{table}[!t]
	\renewcommand{\arraystretch}{1}
	\caption{Architectures for BSNet. * denotes the corresponding module of Res2Net in the implementation code (\url{https://github.com/Res2Net}).}
	\begin{center}
	\setlength{\tabcolsep}{7mm}{
	\begin{tabular}{c|c|c}
\hline\hline
Layer                  & Operation                            & Output                 \\ \hline
\multirow{4}{*}{Conv1} & \multicolumn{1}{l|}{Res2Net.conv1}   & \multirow{4}{*}{$X^1$} \\
                       & \multicolumn{1}{l|}{Res2Net.bn1}     &                        \\
                       & \multicolumn{1}{l|}{Res2Net.relu}    &                        \\
                       & \multicolumn{1}{l|}{Res2Net.maxpool} &                        \\ \hline
Conv2                  & Res2Net.layer1*+ RFB                 & $X^2$                  \\ \hline
Conv3                  & Res2Net.layer2*+ RFB                 & $X^3$                  \\ \hline
Conv4                  & Res2Net.layer3*+ RFB                 & $X^4$                  \\ \hline
Conv5                  & Res2Net.layer4*+ RFB                 & $X^5$                  \\ \hline\hline
\end{tabular}}
\end{center}
\label{table:0}
\end{table}

Considering the important role of semantic relations and boundary constraints on the segmentation task, we design the Dual-branch Semantic Enhancement (DSE) module and Mirror-symmetric Boundary Guidance (MBG) module to highlight the global semantics and sharp boundaries during the decoding process.
The high-level features from the last three convolutional blocks contain abundant semantic information and the corresponding low-level features from the first two convolutional blocks contain more detailed boundary information due to higher spatial resolution. 
In order to comprehensively utilize the rich semantic information of the last three convolutional blocks, we design the DSE module in a complementary two-branch way to generate a global semantic mask, and use it for semantic refinement.
It is well known that clear boundaries are essential for diagnosis, so making full use of the boundary information is the key to obtaining competitive segmentation results. Thus, in addition to the commonly used boundary supervision constraints, we also incorporate the boundary information into the decoding process through the designed MBG module to achieve more in-depth and comprehensive boundary optimization and guidance. 
To strengthen ability to express boundary information, the features of $X^2$ are fed to a total of four convolutional layers to extract boundary information, thereby generating a one-channel mask called boundary map. At the same time, explicit supervision is used between the generated boundary map and the boundary ground truth obtained by the boundary extractor (\eg, Canny) to guarantee the effectiveness of learning.

Finally, we utilize the features obtained by the third decoder layer to generate the final prediction of lung infection regions through additional Sigmoid activation function.

\subsection{Dual-branch Semantic Enhancement Module}
Due to the overwhelming background context redundancies and scattered distribution of the infection regions in the chest CT images, it is difficult to segment COVID-19 lung infection with accurate location and complete structure.
For this problem, resorting to semantic comprehension is a feasible solution. 
The high-level features have been proven to be rich in semantic information, which can construct relationship not only between different scattered regions, but also between complex background and infected regions. Thus, we propose a DSE module to aggregate high-level features in a complementary dual-branch strategy, thereby generating a semantic attention mask $A_{SE}$ to highlight the important regions.    

Different from the existing methods \cite{infnet,AnamNet}, we start with different spatial resolutions and information contents of different high-level features, and design a dual-branch structure, as shown in the lower-right corner of Figure \ref{fig2}.     
On one hand, the top-level features contain the richest channel semantic features, but their spatial resolution is the lowest. Therefore, the most intuitive way of upsampling and filtering is adopted to maintain the pure top-level semantic information. We directly up-sample $X^5$ four times to retain the pure highest-level semantic information. Referring to \cite{crm2022rsi,crm2022tcsvt}, we then employ spatial attention mechanism to obtain attention map $A_t$.
On the other hand, we design a progressive multi-scale fusion strategy, taking into account the information of the three high-level features at the same time, and the final resolution is unified on the scale of the third layer. In this way, the complementary relationship between different high-level features can be learned from a more comprehensive perspective and the spatial resolution sampling distortion can be alleviated. We first implement dilated convolution layers with different dilated rates on $X^3$, $X^4$, and $X^5$ respectively. Then, we progressively up-sample $X^5$ and fuse the deep features with shallow features to achieve adequate high-level features aggregation. The fused features are transformed to attention map $A_p$ by a convolutional layer.

Our final attention map $A_{SE}$ is obtained by adding two attention maps $A_t$ and $A_p$ generated by two complementary attention calculation methods.             

\emph{(1) Top-level Semantic Preservation.}
In this process, we only handle the top-level semantic features. First, we restore the spatial resolution of the features $X^5$ to the resolution of the third-layer features through a $4\times$ upsampling operation. However, the top-level features still contain a lot of redundant information. Therefore, we employ the spatial attention mechanism \cite{cbam} to determine the most important locations in the features and obtain the semantic preservation attention map $A_t$. 
In order to achieve spatial attention, we utilize the average-pooling and max-pooling on upsampled $X^5$ respectively to form two one-channel maps and then concatenate them along the channel axis, thereby generating a two-channel descriptor $\Gamma^s\in{\mathbb{R}^{ H \times W\times 2}}$:
\begin{equation}
	\Gamma^s=concat(avepool(up_4(X^5)),maxpool(up_4(X^5))),
\end{equation}
where $concat(\cdot)$ represents the feature concatenation along channel axis, $up_4(\cdot)$ is the $4\times$ spatial upsampling, $avepool(\cdot)$ and $maxpool(\cdot)$ are the average-pooling and max-pooling, respectively.
Then, the convolution layer with filter sizes of $3 \times 3$ is applied to transform a two-channel descriptor into one 2D attention map $A_t\in{\mathbb{R}^{H \times W}}$:
\begin{equation}
	A_{t}=\sigma(conv_{3\times3}(\Gamma^s;\hat{\theta}_{3\times3})),
\end{equation}
where $\sigma$ denotes the sigmoid function, $conv_{n \times n}$ represents a convolution operation with the filter size of $n \times n$, and $\hat{\theta}_{n \times n}$ is the learnable parameters of the corresponding convolution operation.

\emph{(2) Progressive Semantic Integration.} 
Although we regard the third, fourth, and fifth layers as high-level feature layers, the features they extract are still different. Therefore, in order to obtain more comprehensive semantic information, effectively fusing them is a reasonable solution. In addition, since the size of the infected area varies greatly, in order to allow the model to obtain a robust and stable segmentation result for different regions, we first enhance the features of each layer before fusion to perceive a larger receptive field. 
Concretely, three dilated convolution layers with the dilated rate of $8$, $4$, and $2$ are separately applied to the input features $X^3$, $X^4$, and $X^5$, thereby generating multi-scale features $F_{dc}^3$, $F_{dc}^4$, and $F_{dc}^5$:
\begin{equation}
	\begin{array}{cc}
		& F_{dc}^3=\sigma(conv_{d=8}(X^3;\hat{\omega}_{d=8})), \\
		& F_{dc}^4=\sigma(conv_{d=4}(X^4;\hat{\omega}_{d=4})), \\
		& F_{dc}^5=\sigma(conv_{d=2}(X^5;\hat{\omega}_{d=2})),
	\end{array}
\end{equation}
where $conv_{d=n}$ represents a $3\times3$ convolution operation with the dilated rate of $n$, and $\hat{\omega}_{d=n}$ denotes the learnable parameters. 

Then, in order to reduce the resolution blur distortion caused by upsampling as much as possible in the fusion process, we adopt a progressively fusion strategy.
We first concatenate the upsampled features $F_{d}^5$ with $F_{d}^4$ and employ a $3\times3$ convolution layer to generate the fusion features $F_{4,5}$, which has the same spatial resolution with $X^4$. Similarly, the fused features $F_{4,5}$ are further upsampled and combined with features $F^3_d$, thereby obtaining the global semantic features $F_{3,4,5}$. After that, we employ a convolution layer with filter size of $1\times1$ to generate the semantic integration attention map $A_{p}$. Formally, the above fusion processes can be formulated as:
\begin{equation}
    F_{4,5}=conv_{3\times3}(concat(up_2(F_{d}^5), F_{d}^4)),
\end{equation}
\begin{equation}
    F_{3,4,5}=conv_{3\times3}(concat(up_2(F_{4,5}), F_{d}^3)),
\end{equation}
\begin{equation}
    A_p=\sigma(conv_{1\times1}(F_{3,4,5})),
\end{equation}
where $up_2(\cdot)$ is the $\times2$ spatial upsampling.

Finally, these two attention maps are aggregated to produce the final semantic enhancement attention map $A_{SE}$, which is computed as:
\begin{equation}
	A_{SE}=\frac{1}{2}(A_{t} \oplus A_{p}),
\end{equation}
where $\oplus$ represents the element-wise summation. The attention map $A_{SE}$ is used to further refine decoding process after the MBG module guidance of each stage.

\begin{figure}[!t]
\centerline{\includegraphics[width=0.5\textwidth] {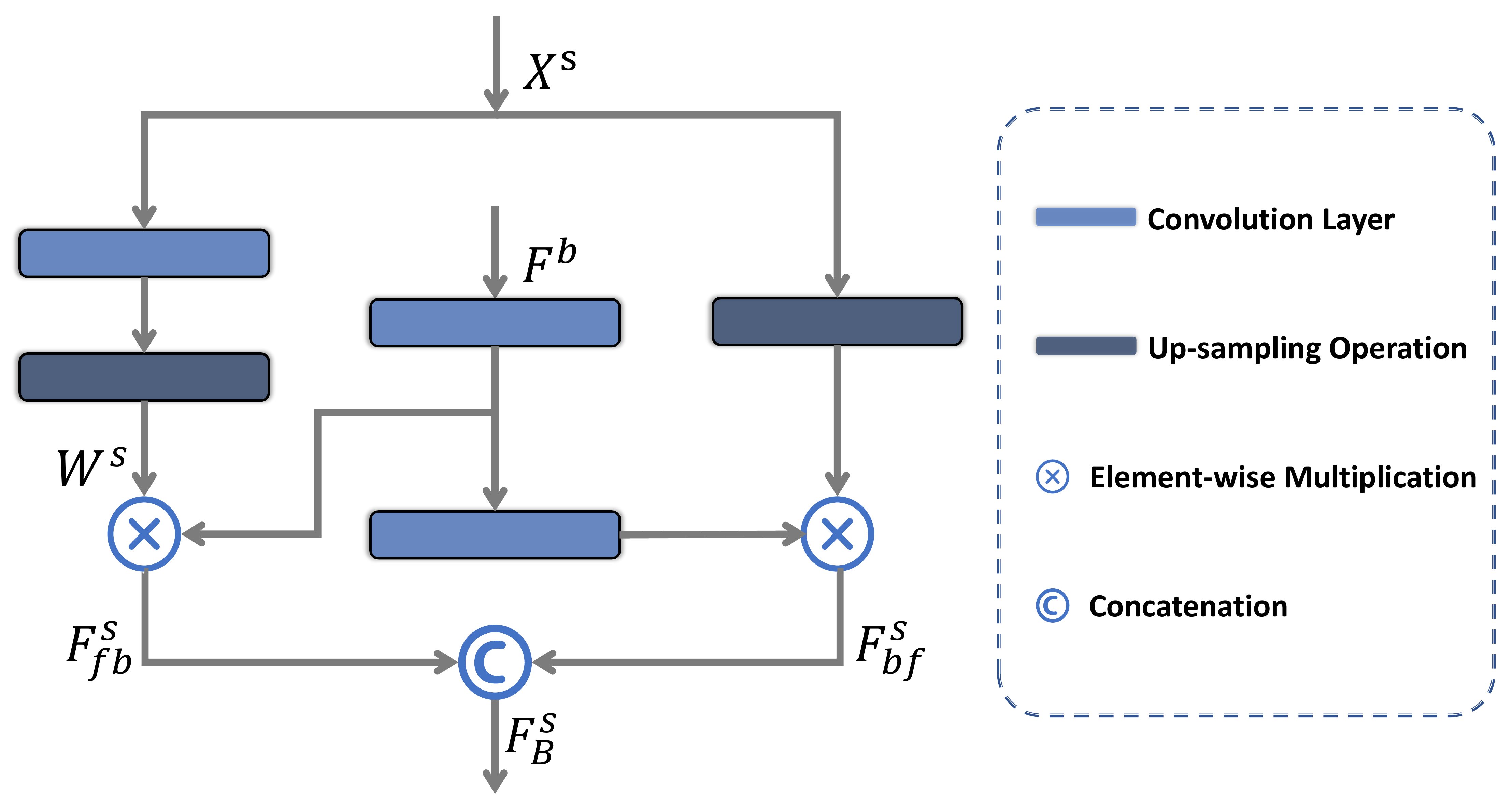}}
\caption{Illustration of the proposed MBG module.}
\label{fig3}
\end{figure}

\subsection{Mirror-symmetric Boundary Guidance Module}
As we all know, low-level features include rich detailed information (\eg, boundaries) with a larger spatial resolution, which are conducive to refine boundaries of the lesion regions accurately. In order to highlight the important boundaries in the feature decoding, we design a MBG module to introduce the boundary guidance of low-level features from the perspective of feature integration, as shown in Figure \ref{fig3}.
Instead of simply combining the features through concatenation or addition operation, the MBG module is designed as a mirror-symmetric structure to combine the corresponding encoder features and the boundary information from the second layer of encoder stage. 
In other words, the boundary features $X^2$ (also denoted as $F_b$) and the corresponding encoder features $X^s$ are worked as the basic features of each other, and other features are used for guidance. 
In the right branch generating $F_{fb}^s$, we modify the encoder features by using the mask derived from boundary features, aiming to reinforce important boundary locations in the encoder features. However, compared to features from the last three convolutional blocks ($X^3$, $X^4$, and $X^5$), $F_b$ may contain much redundant information. Therefore, we generate the corresponding high-level mask to refine the boundary features in the mirror-symmetric left branch. Finally, the two refined features are concatenated to form the final boundary-guided feature output. 
This mirror-symmetrical strategy can ensure the complementarity and sufficiency of feature learning, so as to maximize the interference suppression capability of the multiplication fusion.

Specifically, one is to input the corresponding encoder features $X^s$ and generate a mask $W^s$, thereby refining the boundary features $F_b$. 
For compressing the boundary features $F_b$ (\ie, the features $X^2$ from the second block of encoder stage) down to the same number of channels as the encoder features $X^s~(s\in{\{3,4,5\}})$, we first feed $F_b$ into a $1\times1$ convolution layer. Then, a $3\times 3$ convolution layer with upsampling operation is performed on $X^s$ to obtain a mask $W^s$. Further, we multiply the mask $W^s$ to the channel-suppressed boundary features ${F}_{b}$:
\begin{equation}
    F_{fb}^s = \delta(W^s \odot conv_{1\times1}({F}_{b})),
\end{equation}
where $W^s = up_n(conv_{3\times3}(X^s))$, $n=2^{s-2}$ is the upsampling scale factor, $\odot$ denotes the element-wise multiplication, $\delta$ denotes the ReLU activation function, and $s$ indexes the feature level.

The other is a mirror symmetry method with the first one, which modifies the encoder features by using the mask derived from boundary features.
Specifically, we obtain the boundary mask by performing an additional $3\times 3$ convolution operation on the boundary features ${F}_{b}$, and multiply it by the upsampled features $X^s$:
\begin{equation}
    F_{bf}^s = up_n(X_s) \odot conv_{3\times3}(conv_{1\times1}({F}_{b})),
\end{equation}
where $up(\cdot)$ upsamples $X_s$ to the same resolution of ${F}_{b}$.
Then, we concatenate $F_{fb}^s$ and $F_{bf}^s$ to obtain the output of MBG module \ie, $F_B^s$. 
Each MBG module is followed by a channel-wise concatenation to integrate the upsampled decoder features from the previous stage and boundary-guided features. After that, the features after concatenation are fed into a $3\times 3$ convolution layer for compressing to the original channel number. Finally, a skip connection is employed to generate the current decoder features $F_{d}^s$:
\begin{equation}
    F_{d}^s = F_{d}^{s+1} \oplus conv_{3 \times 3} (concat(F_B^s,conv_{3 \times 3}(up_2(F_d^{s+1})))),
\end{equation}
where $s\in{\{5,4,3\}}$ indexes the decoder stage. Note that, for the calculation of the top-level decoder features $F_d^5$, since there is no previous decoder layer, we directly use the semantic enhancement attention map instead, \ie, $F_d^6=A_{SE}$.

\subsection{Loss Function}
We design a hierarchical loss function on the side outputs of different scales (\ie, $S_3$, $S_4$, and $S_5$) and semantic attention map $A_{SE}$ by weighted IoU loss and weighted BCE loss. Following \cite{l1-qin2019basnet}, \cite{l2-wei2020f3net}, compared with the traditional IoU loss and BCE loss, the weights in the weighted IoU/BCE loss pay more attention on hard pixels and assign larger weights to them. In addition, local structure information is encoded into the weighted BCE loss, which may help the model focus on a larger receptive field rather than on a single pixel. Specifically, the weighted BCE loss and weighted IoU loss are defined as:
\begin{equation}
\resizebox{.8\hsize}{!}{$l_{wbce}^k=-\frac{\sum\limits_{i=1}^H\sum\limits_{j=1}^W(1+\gamma\alpha_{ij})\sum\limits_{l=0}^1\textbf{1}(gt_{ij}^k=l)log\textbf{Pr}(p_{ij}^k=l\mid\Psi)}{\sum\limits_{i=1}^H\sum\limits_{j=1}^W\gamma\alpha_{ij}}$},
\end{equation}
\begin{equation}
\resizebox{.8\hsize}{!}{$l_{wiou}^k=1-\frac{\sum\limits_{i=1}^H\sum\limits_{j=1}^W(gt_{ij}^k\cdot p_{ij}^k)\cdot(1+\gamma\alpha_{ij}^k)}{\sum\limits_{i=1}^H\sum\limits_{j=1}^W(gt_{ij}^k+p_{ij}^k-gt_{ij}^k \cdot p_{ij}^k) \cdot (1+\gamma\alpha_{ij}^k)}$},
\end{equation}
where $W$ and $H$ are the width and height of the input image, $\textbf{1}(\cdot)$ is the indicator function, $l=\{0,1\}$ indicates two kinds of labels, $\gamma$ is a hyper parameter, $p_{ij}^k$ and $gt_{ij}^k$ are the prediction and ground truth of the pixel at location $(i,j)$ in the image $k$, $k=\{S_3,S_4,S_5,A_{SE}\}$, $\Psi$ represents all the parameters of the model, $\textbf{Pr}(p_{ij}^k=l\mid\Psi)$ denotes the predicted probability, and $\alpha_{ij}^k=\left|\frac{\sum_{(m,n)\in{A_{ij}}}gt_{mn}^k}{\sum_{(m,n)\in{A_{ij}}}1}-gt_{ij}^k\right|$ is a pixel importance indicator, which is calculated by the difference between the center pixel and its surrounding pixel set $A_{ij}$.
For simplicity, we do not distinguish the importance of these two losses in the final loss function, and the weighting coefficients of weighted BCE loss and weighted IoU loss are both set to 1. 

Moreover, we use the standard binary cross-entropy on the boundary map as the boundary-aware loss function. Therefore, the total loss can be defined as:
\begin{equation}
\iota_{total} = l^b_{bce}+\sum_{k=\{S_3,S_4,S_5,A_{SE}\}}(l^k_{wiou}+l^k_{wbce}),
\end{equation}
where $l^b_{bce}$ is the boundary loss using the standard binary cross-entropy. 

\section{Experiments} \label{sec4}

\subsection{Benchmark Dataset and Evaluation Metrics}
\textbf{Benchmark Dataset.} Research shows that there are currently fewer public COVID-19 lung CT datasets used for infection segmentation. To be able to train the model better, we merge the two publicly available CT-based segmentation datasets \cite{segdata1,segdata2} to obtain 1018 CT images, which are further divided into 718 training images and 300 testing images. The design of our method requires boundary supervision information, so the Canny operator is used to extract the boundaries of the infection mask. Each CT slice contains the original image, the corresponding infection mask, and the corresponding infection boundary.


\textbf{Evaluation Metrics.} We use seven metrics for quantitative evaluation, \ie, Dice Similarity Coefficient (DSC) \cite{2020Lung}, Sensitivity (Sen.) \cite{2020Large}, Structure Measure ($S_\alpha$) \cite{S}, Enhance-alignment Measure ($E_\phi$) \cite{E}, Mean Absolute Error (MAE) \cite{crm2019tcsvt,crmJEI,crmSPIC,crm2019tip,crm2020going}, Precision (Prec.) \cite{crmICME,crm2018tip,crm2019tc,crm2019tmm}, and Hausdorff Distance (HD) \cite{huttenlocher1993comparing}. The Prec. is the proportion of positive samples in the samples that are predicted to be positive, and the Sen. assesses the ratio of correctly identified positive cases to all positive cases.

The DSC is used to evaluate the overlap ratio between the predicted segmentation map $S_p$ and the corresponding ground truth $G$, which is calculated by:
\begin{equation}
	\text{DSC}=\frac{2\cdot|G \cap S_p|}{|G| + |S_p|}.
\end{equation}

The $S_\alpha$ measures the structural similarity between the segmentation map $S_p$ and the ground truth $G$:
\begin{equation}
	S_\alpha=(1-\alpha)\cdot S_o+\alpha \cdot S_r,
\end{equation}
where $\alpha$ is set to 0.5 for assigning equal contribution to both region similarity $S_r$ and object similarity $S_o$.

\begin{figure}[!t]
\centerline{\includegraphics[width=9cm,height=4cm] {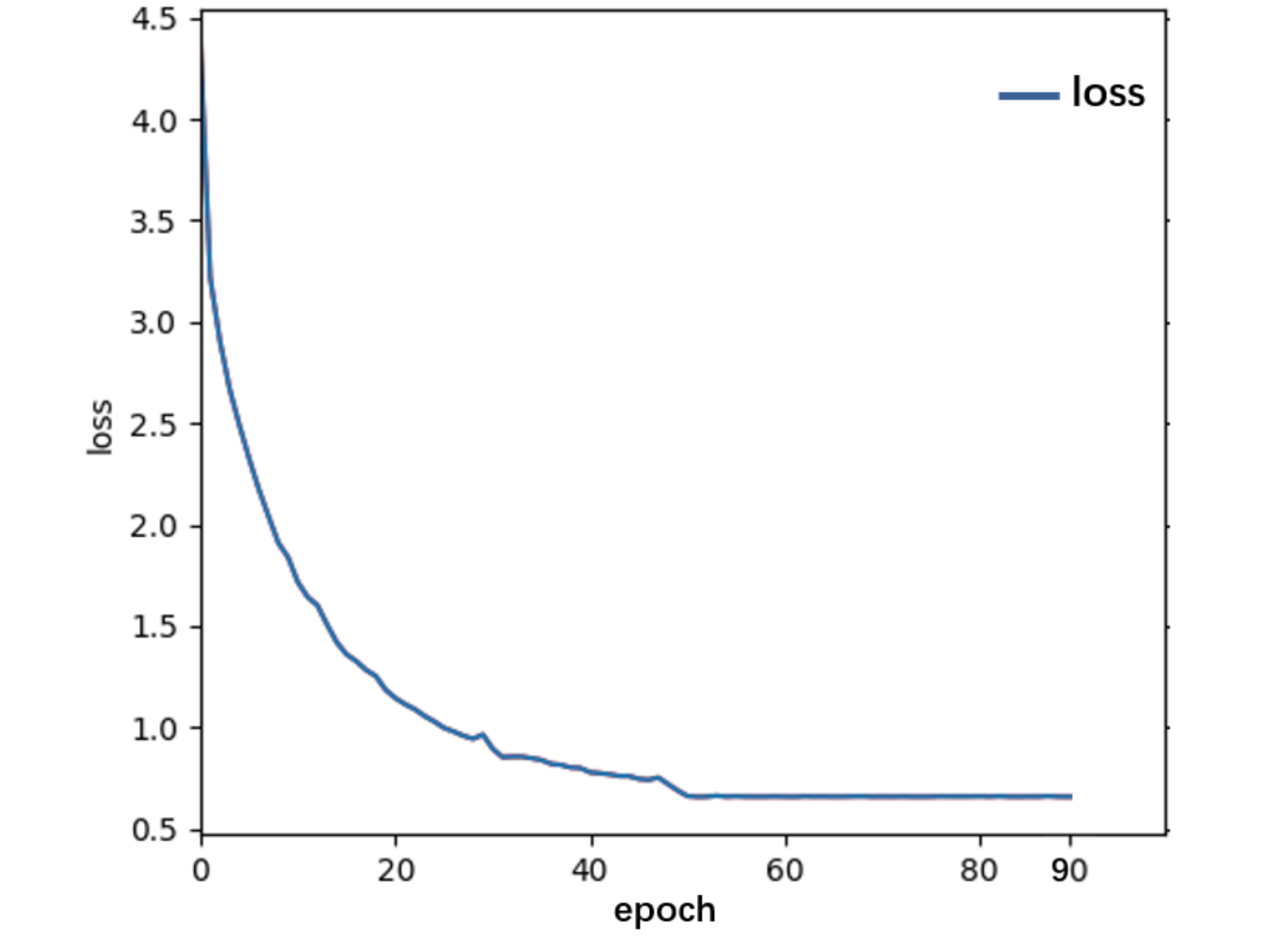}}
\caption{The training loss curve.}
\label{loss}
\end{figure}

The $E_\phi$ is used to evaluate both local and global similarity between two binary maps, which is formulated as:
\begin{equation}
	E_\phi=\frac{1}{w \times h}\sum_x^w\sum_y^h\phi(S_p(x,y),G(x,y)),
\end{equation}
where $w$ and $h$ are the width and height of the image, $(x, y)$ denotes the coordinate of each pixel in $S_p$ and $G$, and $\phi$ is the enhanced alignment matrix.

\begin{figure*}[!t]
\centerline{\includegraphics[width=18cm,height=10cm]{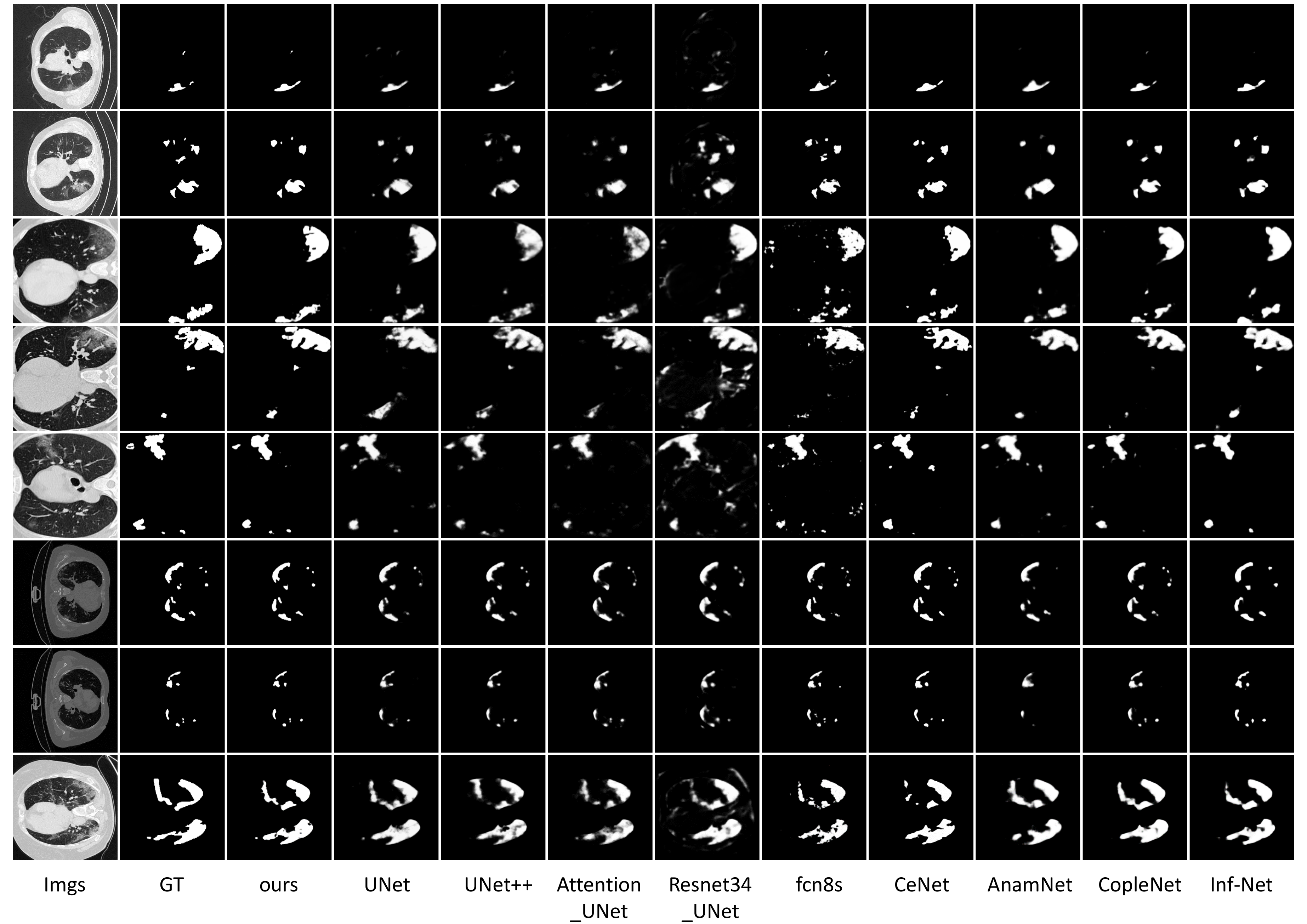}}
\caption{Visual comparisons of different methods.}
\label{Visual comparisons}
\end{figure*}

The MAE measures the pixel-level error between the prediction map $S_p$ and the ground truth $G$, which is defined as:
\begin{equation}
\text{MAE}=\frac{1}{w \times h}\sum_x^w\sum_y^h|S_p(x,y)-G(x,y))|.
\end{equation}

The HD explicitly measures the boundary performance, which is defined as:
\begin{equation}
	\text{HD}=\max(\max\limits_{p \in{B}}\{\min\limits_{q \in{G_B}}\Vert p-q\Vert\},\max\limits_{q \in{G_B}}\{\min\limits_{p \in{B}}\Vert q-p\Vert\}),
\end{equation}
where $p$ and $q$ represent the pixel in the boundary prediction set $B$ and boundary ground truth  set $G_B$, respectively.

Among these indicators, Sen. and $S_\alpha$ can reflect the segmentation integrity, HD measures the boundary effects, and the DSC, Prec., $E_\phi$ and MAE can evaluate the overall performance.
Moreover, in addition to the MAE and HD, the larger the value, the better the performance.

\subsection{Implementation Details}

We implement the proposed model via the PyTorch toolbox and train it on an RTX 2080Ti GPU in an end-to-end manner. We also implement our network by using the MindSpore Lite tool\footnote{\url{https://www.mindspore.cn/}}. Referring to Inf-Net \cite{infnet}, the Res2Net-50 \cite{Res2Net} pretrained on ImageNet \cite{imagenet} is employed as the backbone feature extractor in the experiment. 
In addition to the reason for fair comparison, the reason why we choose Res2Net-50 as the backbone network also benefits from its own advantages. First, compared to the classical ResNet \cite{resnet} and VGG \cite{vgg}, the Res2Net can construct hierarchical residual-like connections within one single residual block, represent multi-scale features at a granular level, and increase the range of receptive fields for each network layer. In addition, the Res2Net consumes less parameters and computing resources, which is also very friendly to improve the real-time efficiency of our model. 
Due to limited computing resources, all input images are resized to $352\times 352$, and a multi-scale training strategy \cite{strategy} is used to train the network. Our BSNet is trained by using the Adam optimizer \cite{adam} for 90 epochs, the batch size and learning rate are set to 8 and $1e^{-4}$ respectively. We choose the model according to the determined epoch number. As can be seen from the training curve shown in Figure \ref{loss}, our network can converge after training to 90 epochs.

\subsection{Comparison with the State-of-the-art Methods}
In order to demonstrate the effectiveness of BSNet, we compare it with eleven state-of-the-art methods, including UNet \cite{unet}, UNet++ \cite{unet++}, Attention\_UNet \cite{attentionunet}, ResNet34\_Unet \cite{unet}, CeNet \cite{Cenet}, fcn8s \cite{fcn8s}, CopleNet \cite{COPLENet}, JCS \cite{wu2021jcs}, FFR \cite{wang2021focus}, AnamNet \cite{AnamNet}, and Inf-Net \cite{infnet}. To ensure the fairness of the experiment, all state-of-the-art methods are retrained on the same dataset as our BSNet under the default parameters.     

\textbf{Qualitative Comparison.}
Figure \ref{Visual comparisons} shows the visual comparison results of different methods. We can see that our method more accurately and completely locates the COVID-19 lung infection regions than other competing methods. 
On the whole, the classic segmentation methods (\eg, UNet \cite{unet}, UNet++ \cite{unet++}, Attention\_UNet \cite{attentionunet}, and ResNet34\_UNet \cite{unet}) tend to have weak background interference suppression capabilities, leading to erroneous prediction results. 
By contrast, the existing COVID-19 segmentation methods (\eg, Inf-Net \cite{infnet}, CopleNet \cite{COPLENet}, and AnamNet \cite{AnamNet}) achieve better detection results, but these methods fail to completely suppress the background, thereby leading to false detection and missing detection to some extent. For example, the background regions in the middle of the fifth image are not effectively suppressed, and there is the missing detection phenomenon in the lower area of the image.
However, our proposed method exhibits stronger advantages in terms of accurate positioning, background suppression, and detection integrity. In addition, the lower right area of the last image, only our method can locate the infected regions clearly, accurately, and completely. In general, our method has a more complete structure and clearer boundaries, which benefits from the full use of high-level semantic information and edge information for modeling.

\begin{table}
	\centering
	\caption
	{Quantitative comparisons with different methods. “CLA” denotes the classical segmentation model, and “COV” represents the segmentation model for COVID-19. The best and second best performance are are bolded and underlined. $\uparrow \& \downarrow$ denote larger and smaller is better, respectively.}
	\begin{center}
	\renewcommand{\arraystretch}{1.1}
	\setlength{\tabcolsep}{0.8mm}{
\begin{tabular}{c|c||ccccccc}
\hline\hline
Method         &Type & DSC$\uparrow$  & Sen.$\uparrow$  & $S_\alpha$$\uparrow$  & $E_\phi$$\uparrow$  & MAE$\downarrow$   & Prec.$\uparrow$ & HD$\downarrow$ \\\hline
UNet            &CLA & 0.777 & 0.814 & 0.862 & 0.917 & 0.020 & 0.804 & 27.219   \\
UNet++         &CLA & 0.771 & 0.780 & 0.867 & 0.906 & 0.021 & 0.836 & 26.081   \\
Attention\_UNet &CLA & 0.746 & 0.768 & 0.853 & 0.886 & 0.021 & 0.818 & 29.718  \\
ResNet34\_UNet  &CLA & 0.720 & 0.836 & 0.812 & 0.873 & 0.030 & 0.702 & 42.140   \\
fcn8s          &CLA & 0.800 & 0.791 & 0.855 & 0.949 & 0.020 & 0.839 & 27.286  \\
CeNet          &CLA & 0.818 & 0.824 & 0.854 & 0.960 & 0.017 & 0.834 & 44.589  \\\hline
CopleNet       &COV & 0.816 & 0.821 & 0.874 & 0.944 & 0.016 & 0.850 & 25.908  \\

AnamNet        &COV & 0.775 & 0.776 & 0.856 & 0.920 & 0.021 & 0.831 & 35.401  \\

JCS       & COV & 0.836 & 0.835 & 0.869 & 0.965 & 0.017 &  \underline{0.855} & 24.559 \\

FFR       &COV & \underline{0.839} & 0.841 & 0.869 & \underline{0.971} & \underline{0.015} & 0.852  & \underline{19.643}  \\
Inf-Net        &COV & 0.828 & \underline{0.846} & \underline{0.877} & 0.963 & 0.016 & 0.831 & 24.403  \\
ours           &COV & \textbf{0.851} & \textbf{0.849} & \textbf{0.884} & \textbf{0.973} & \textbf{0.014} & \textbf{0.867} & \textbf{19.462} \\\hline\hline
\end{tabular}
			}
	\end{center}
	\label{table:1}
\end{table}

\textbf{Quantitative Comparison.} 
The numerical indexes, including Dice Similarity Coefficient (DSC), Sensitivity (Sen.), Precision (Prec.), Structure Measure ($S_\alpha$), Enhance-alignment Measure ($E_\phi$), Mean Absolute Error (MAE), and Hausdorff Distance (HD), are reported in Table \ref{table:1}. It is evident that our model achieves competitive performance across different metrics. To be speciﬁc, our method achieves the best performance in terms of all measures on the merged dataset.

Compared with the classical deep learning-based segmentation methods for natural imsges (\ie, UNet \cite{unet}, UNet++ \cite{unet++}, Attention\_UNet \cite{attentionunet}, and ResNet34\_UNet \cite{unet}), most of the segmentation methods speciﬁcally designed for COVID-19 exhibit the superior performance, demonstrating the particularity and challenges of COVID-19 infection segmentation. 
It's worth mentioning that AnamNet is an embedding-based lightweight network with about one-sixth parameter consumption of the classic U-Net, which is why its performance is not as good as some classic medical segmentation algorithms.
In summary, due to the delicately designed modules, our BSNet ranks first in all evaluation metrics. 
For example, compared with the classic UNet \cite{unet} method, the percentage gain of our method reaches 9.52\% in terms of the DSC, and 30.00\% in terms of the MAE score. 
Compared with the \emph{second best} method, the percentage gain reaches 1.43\% in terms of the DSC, 1.40\% in terms of Prec., and 6.67\% in terms of MAE, respectively. 
In addition, similar to the visualization results, some metrics, such as Sen. and  $S_\alpha$, reflecting the detection completeness also demonstrate the advantages of our method.
In terms of these two indicators, our method achieves the best results compared to other competitors. Specifically, compared to the \emph{second best} method, the percentage gain reaches 0.35\% in terms of the Sen., and 0.80\% in terms of $S_\alpha$. 
Likewise, our method achieves the best performance in boundary effect evaluation by using the HD score. For example, our method wins a minimum percentage gain of 0.9\%, and a maximum percentage gain of 56.4\% against the comparison methods.
All these clearly demonstrate the superior performance of the proposed model in COVID-19 lung infection segmentation.

\begin{figure}[!t]
\centerline{\includegraphics[width=0.5\textwidth] {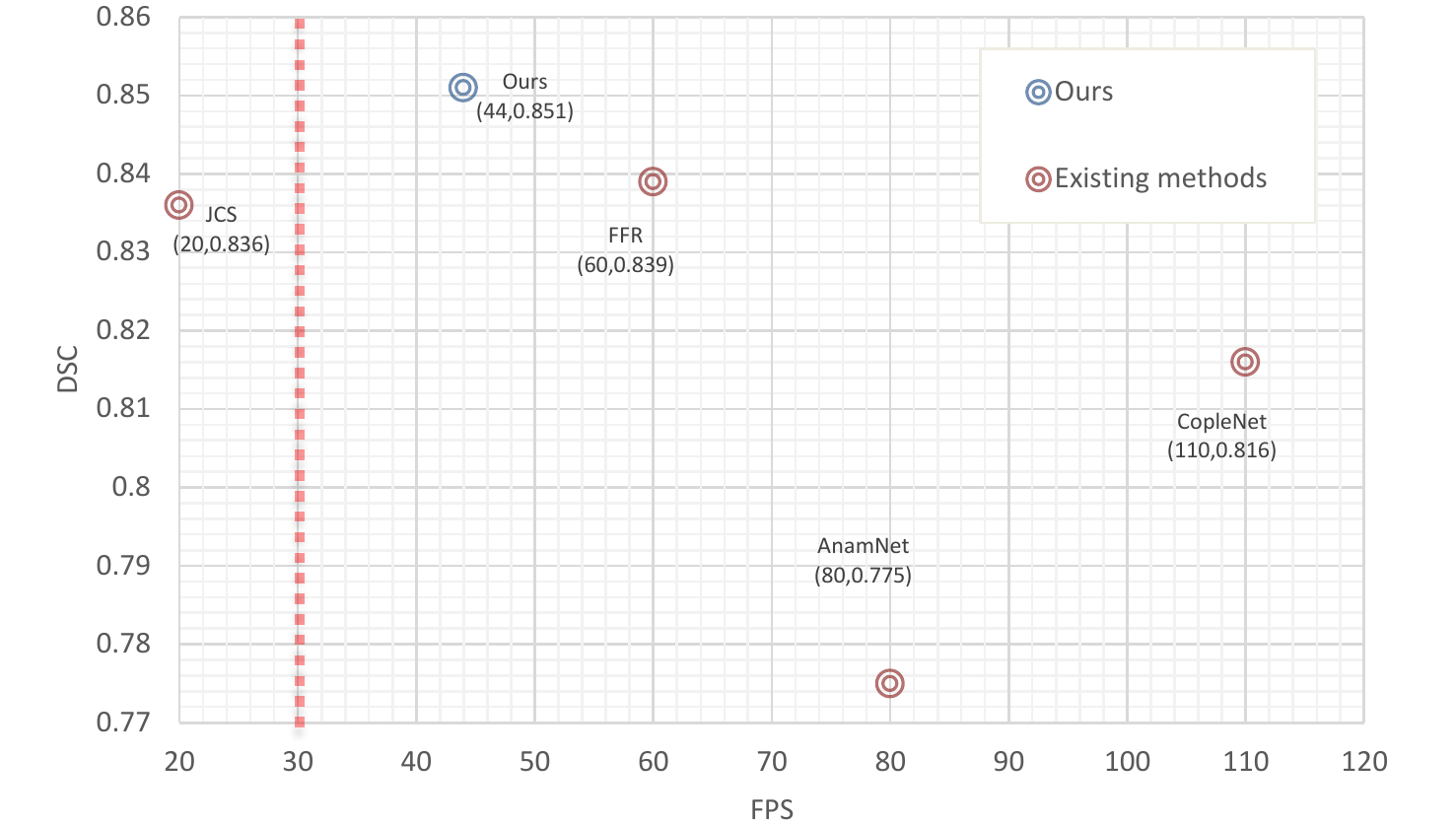}}
\caption{The FPS-DSC map for different methods.}
\label{speed}
\end{figure}

In order to make the inference speed comparison with other methods succinctly and clearly, we provide the FPS-DSC map in Figure \ref{speed}. Generally speaking, more than 30 FPS can be considered real-time, and our model reaches 44 FPS, which meets the real-time requirement. While some algorithms are fast at inference (\eg, AnamNet \cite{AnamNet}, CopleNet \cite{COPLENet}), their performance is not as good as ours. In other words, our method strikes a trade-off between performance and efficiency.

\subsection{Ablation Study}
\subsubsection{Validation of key modules} We conduct ablation experiments to verify the effectiveness of the each key module of our proposed model, including the MBG module and DSE module. The quantitative results are shown in Table \ref{table:2}.



\begin{figure}[!t]
\centerline{\includegraphics[width=0.5\textwidth] {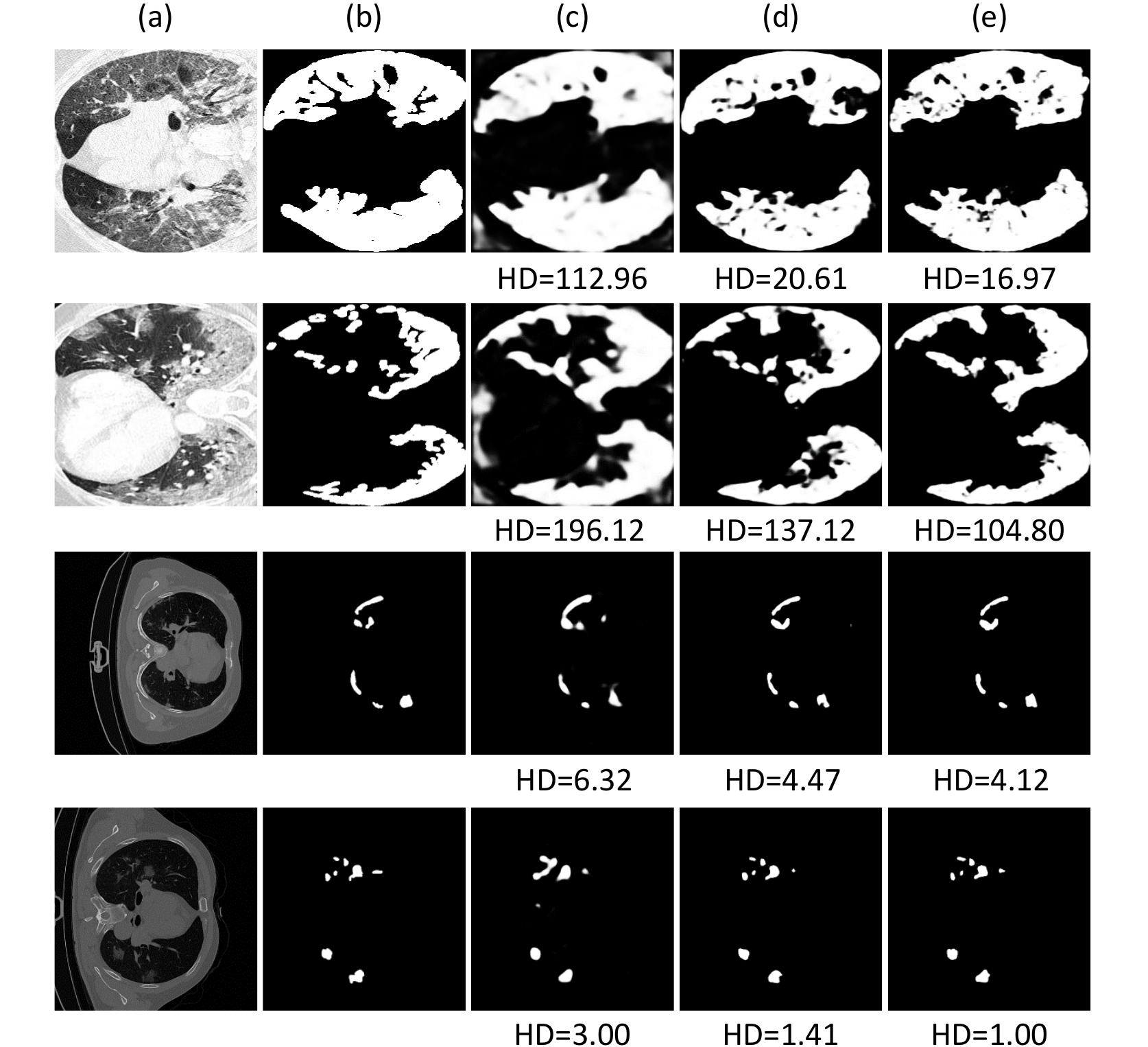}}
\caption{Visual comparisons of BSNet variants equipped with different modules, and the HD scores are also reported below each image. (a) Images. (b) Ground truth. (c) Baseline. (d) Baseline + DSE. (e) Baseline + DSE + MBG (Full Model).}
\label{fig4}
\end{figure}

In model 2, the DSE module is first added to the baseline to deal with background context redundancies and scattered distribution of the infection regions in the chest CT image. Compared to baseline (model 1), the high-level features are not directly used as global information for guidance, but refined by DSE module. Compared with the baseline (model 1), all indicators in model 2 increase obviously, especially for the DSC and MAE. Specifically, the DSC is boosted from 0.764 to 0.838 with the percentage gain of 9.69\%, and the MAE is improved from 0.024 to 0.017 with the percentage gain of 29.17\%. It further shows that the DSE proposed in this paper can more effectively use high-level semantic information to provide the global guidance. We present the qualitative comparison results in Figure \ref{fig4}. Compared with the Baseline model shown in Figure \ref{fig4}(c), it can be found that some irreverent interferences and complex backgrounds around the infection region are well suppressed by introducing the DSE module (\eg, the leftmost region in the second image and the small upper right region in the third image).

\begin{table}
	\centering
	\caption{Quantitative evaluation of ablation study. $\uparrow \& \downarrow$ denote larger and smaller is better, respectively.}
	\begin{center}
		\renewcommand{\arraystretch}{1}
		\setlength{\tabcolsep}{0.5mm}{
\begin{tabular}{c|ccc||ccccccc}
\hline\hline
ID & Baseline   & DSE        & MBG        & DSC$\uparrow$    & Sen.$\uparrow$   & $S_\alpha$$\uparrow$ & $E_\phi$$\uparrow$ & MAE$\downarrow$  & Prec.$\uparrow$   &HD$\downarrow$  \\ \hline
1  & \checkmark &            &            & $0.764$          & $0.834$          & $0.826$              & $0.927$            & $0.024$          & $0.743$        & 45.919               \\
2  & \checkmark & \checkmark &            & $0.838$          & $0.837$          & $0.876$              & $0.969$            & $0.017$          & $0.855$        & 20.775           \\
3  & \checkmark & \checkmark & \checkmark & $\textbf{0.851}$ & $\textbf{0.849}$ & $\textbf{0.884}$     & $\textbf{0.973}$   & $\textbf{0.014}$ & $\textbf{0.867}$  & 19.462 \\ \hline\hline
\end{tabular}
			}
	\end{center}
	\label{table:2}
\end{table}

We discuss the impact of the MBG module on the entire network and design the ablation experiments. The model 3 (full model) is to verify the importance of the boundary information, that is, add the MBG module to aforementioned architecture. It can be seen from Table \ref{table:2} that the introduction of boundary information is effective and can bring performance gains. Compared with the model 2, the DSC and Sen. scores of model 3 are improved by 1.30\% and 1.20\%, respectively. For the visualization evaluation of the boundary effect, we can start from the following two aspects: (1) Boundary accuracy means that the detected boundary structure is complete, the location is accurate, and the degree of coincidence with the boundary GT is high. For example, compared with column (d) in Figure \ref{fig4}, the tiny regions in the upper right corner of the first image and the bottom right corner of the third image are all detected by our model after introducing the MBG unit. (2) Boundary sharpness refers to the boundaries of the detection result are sharp, clear and non-blurred, which is of great significance for diagnosis and treatment. For example, the boundaries of the large infected regions in the lower half of the second image are obviously sharper than the results in the column (d) of Figure \ref{fig4}. In order to observe the changes of boundary effects more clearly and intuitively, we provide the HD scores below the visualization results. It can be seen that the boundary effect has been improved after introducing the MBG module. In addition, in order to verify the design of mirror-symmetric structure in MBG module, we design an additional ablation experiment with only the left branch or right branch, as shown in Table \ref{table:3}. We can clearly see that the absence of any branch in the MBG module will lead to inferior performance, which illustrates the necessity of our mirror-symmetric design.

\begin{table}[!t]
    
	\renewcommand{\arraystretch}{1}
	\caption{The performance comparisons without the left or right branch in MBG module.}
	\begin{center}
	\setlength{\tabcolsep}{1.3mm}{
	\begin{tabular}{c|ccccccc}
\hline\hline
      & DSC$\uparrow$  & Sen.$\uparrow$  & $S_\alpha$$\uparrow$  & $E_\phi$$\uparrow$  & MAE$\downarrow$   & Prec.$\uparrow$  &HD$\downarrow$     \\ \hline
MBG       & $0.851$   & $0.849$ & $0.884$ & $0.973$ & $0.014$   & $0.866$   & 19.462\\ 
MBG w/o $F_{fb}^s$     & $0.847$   & $0.842$ & $0.879$ & $0.970$ & $0.014$   & $0.857$   & 26.162\\ 
MBG w/o $F_{bf}^s$ & $0.846$   & $0.847$ & $0.879$ & $0.963$ & $0.016$   & $0.847$  & 23.156 \\\hline\hline
\end{tabular}
	}
\end{center}
\label{table:3}

\end{table}

\begin{table}[!t]
	\renewcommand{\arraystretch}{1}
	\caption{The performance comparisons of different boundary guidance methods.}
	\begin{center}
	\setlength{\tabcolsep}{1mm}{
	\begin{tabular}{c|ccccccc}
\hline\hline
      & DSC$\uparrow$  & Sen.$\uparrow$  & $S_\alpha$$\uparrow$  & $E_\phi$$\uparrow$  & MAE$\downarrow$   & Prec.$\uparrow$ & HD$\downarrow$      \\ \hline
$X^1$; Canny      & $0.846$   & $0.848$ & $0.877$ & $0.972$ & $0.015$   & $0.858$  & 21.445 \\ 
$X^5$; Canny     & $0.714$   & $0.835$ & $0.794$ & $0.909$ & $0.025$   & $0.650$   & 34.579\\ 
\hline
$X^2$; Sobel       & $0.851$   & $0.849$ & $0.884$ & $0.973$ & $0.014$   & $0.866$  & 20.750 \\ 
$X^2$; Roberts     & $0.851$   & $0.850$ & $0.882$ & $0.974$ & $0.014$   & $0.866$  & 22.243 \\ 
\hline
Ours ($X^2$; Canny) & $0.851$   & $0.849$ & $0.884$ & $0.973$ & $0.014$   & $0.867$   & 19.462\\\hline\hline
\end{tabular}
	}
\end{center}
\label{table:guidance}

\end{table}


\subsubsection{Validation of different boundary guidance methods}  
To verify how boundary supervision and boundary features are generated, we design two ablation experiments.

First, the boundary supervision map is directly processed on the binarized segmentation ground truth, which is a very simple computational task. Theoretically, although different boundary extractors result in slightly different boundary ground truths, there is no significant difference in general boundary locations. To verify this, we design an ablation experiment with three boundary GT extraction methods, including Canny operator, Sobel operator, and Roberts operator, as shown in Table \ref{table:guidance}. From it, we can see that there is almost no difference in the final segmentation results obtained by different boundary GT generation algorithms, which confirms our previous conjecture.

Second, as demonstrated in the existing works \cite{infnet}, the low-level features (\eg, $X^1$, $X^2$) have a larger spatial resolution and include rich detailed information such as boundaries, which are conducive to refine boundaries of the lesion regions accurately. In our method, $X^2$ is chosen to provide boundary guidance for the MBG module, because $X^1$ contains a lot of unimportant and indistinguishable information, which is disruptive or burdensome for feature purifying. To this end, we add an ablation study to verify the effect of different boundary features. As shown in Table \ref{table:guidance}, the top-level features $X^5$ containing rich semantic information are the worst, which also verifies the validity and rationality of our use of low-level features as boundary features. Furthermore, the boundary guidance from the features of $X^2$ achieves better performance than using $X^1$, which illustrates the effectiveness of our setup.

    



\subsubsection{Validation of different output stages} 
The final output of the network is derived from features at the third decoding stage, mainly based on the following two points. First, we design the MBG module to make full use of the features of the second encoder layer ($X^2$) to supplement boundary information for high-level encoder features ($X^3$, $X^4$, and $X^5$). Under such a model framework, we do not embed the MBG module in shallow layers, so we also do not perform decoding at these stages.  Second, in order to achieve a real-time inference speed, generating the final map from higher stage with lower resolution will consume fewer computing resources. In order to verify the difference in performance of different decoding stages, we design an ablation experiment. For fair comparison, we implement the same structure as the third decoding stage on the first and second decoding stage, which contains the MBG module, a channel-wise concatenation and a $3\times3$ convolution layer. As shown in Table \ref{table:5}, it can be found that the third decoding layer achieves better performance than using the first and second decoder features. 

\begin{table}[!t]
    
	\renewcommand{\arraystretch}{1}
	\caption{The performance comparisons of outputs from 1st, 2nd, and 3rd decoding stages, where `-DE' means decoding stage.}
	\begin{center}
	\setlength{\tabcolsep}{1.3mm}{
	\begin{tabular}{c|ccccccc}
\hline\hline
      & DSC$\uparrow$  & Sen.$\uparrow$  & $S_\alpha$$\uparrow$  & $E_\phi$$\uparrow$  & MAE$\downarrow$   & Prec.$\uparrow$ & HD$\downarrow$    \\ \hline
1st-DE      & $0.820$   & $0.802$ & $0.859$ & $0.954$ & $0.019$   & $0.864$ & 22.203 \\ 
2nd-DE     & $0.819$   & $0.827$ & $0.864$ & $0.958$ & $0.018$   & $0.836$  & 23.540 \\ 
3rd-DE (Ours) & $0.851$   & $0.849$ & $0.884$ & $0.973$ & $0.014$   & $0.867$  & 19.462 \\\hline\hline
\end{tabular}
	}
\end{center}
\label{table:5}
\end{table}

\section{Conclusion and Future Work} \label{sec5}
This paper proposes a boundary guided semantic learning network for automatically segmenting COVID-19 lung infections from CT images by studying how to capture the infection area from the perspective of semantic relation and boundary guidance. 
The DSE module models semantic relations through complementary dual-branch strategies, and MBG module adopts mirror symmetry structure to ensure the complementarity and sufficiency of feature learning. Experiments show that our BSNet outperforms the state-of-the-art competitors and achieves the real-time effects. 

Although our algorithm achieves a more complete structure and more accurate details, it is very challenging for COVID-19 infection segmentation due to the scattered infected regions over the chest slice. It just so happens that Transformer (\eg, ViT \cite{dosovitskiy2020image} and Swin Transformer \cite{liu2021swin}), which can model long-term dependencies in data through self-attention mechanism, have been widely used in medical image segmentation \cite{wu2022fat,he2022fully}. Therefore, we believe that exploring Transformer-based COVID-19 infection segmentation task is a worthy future research direction. In addition, considering the contradiction between the model and data, on the one hand, global scientific research institutions can be called on to open source relevant data in accordance with relevant regulations. On the other hand, we can try to use domain adaptation \cite{zhang2022unsupervised,lei2021unsupervised} to better transfer the model trained on the normal medical segmentation dataset to the COVID-19 infection segmentation, which technically makes up for the lack of training data.

\par
\ifCLASSOPTIONcaptionsoff
  \newpage
\fi
{
\bibliographystyle{IEEEtran}
\bibliography{ref}
}

\end{document}